\documentclass[sensors,accept,moreauthors,pdftex,10pt,a4paper]{mdpi}
\firstpage{1}
\articlenumber{}
\doinum{}
\pubvolume{}
\pubyear{}
\copyrightyear{}
\externaleditor{}
\history{}
\usepackage{epsfig}
\usepackage{color}
\usepackage{enumerate}
\usepackage{easylist}
\usepackage[table]{xcolor}
\usepackage{amsmath,graphicx}
\usepackage{subfigure}
\usepackage{multirow}
\usepackage{float}
\usepackage{longtable}
\usepackage{amsthm}
\usepackage{amssymb }
\usepackage{amsmath}
\usepackage{graphicx}
\usepackage{mathrsfs}
\usepackage{amsfonts}
\usepackage{longtable}
\usepackage{multirow}
\usepackage{slashbox}
\usepackage[T1]{fontenc}
\usepackage{amsmath}
\usepackage[cmintegrals]{newtxmath}
\usepackage{bm}
\usepackage{array}
\usepackage{rotating}

\def\NH{N_h}

\def\y{\bm{y}}

\def\w{\bm{w}}
\def\Y{Y}

\def\X{\bm{X}}

\def\DFS{\text{DF-SSmVEP}} 
\def\DCCA{\text{BCCA}}
\def\Tar{^{(\text{T})}}
\def\NT{N^{(\text{T})}}
 \theoremstyle{mdpi}
 \newcounter{thm}
 \newcounter{ex}
 \newcounter{re}
 \theoremstyle{mdpidefinition}

\Title{$\DFS$: Dual Frequency Aggregated Steady-State Motion Visual  Evoked Potential Design with Bifold Canonical Correlation Analysis} 

\Author{Raika Karimi$^{1}$, Arash Mohammadi $^{1, 2}$, Amir Asif $^{3}$, and Habib Benali $^{1}$}
\AuthorNames{Raika Karimi, Arash Mohammadi, Amir Asif, and Habib Benali}

\address{%
$^{1}$  Department of Electrical and Computer Engineering, Concordia University, Montreal, Canada.\\
$^{2}$ Concordia Institute for Information System Engineering, Concordia University, Montreal, Canada; arash.mohammadi@concordia.ca\\
$^{3}$ Department of Electrical Engineering and Computer Science, York University, Toronto, Canada.}

\corres{Correspondence: arash.mohamad@concordia.ca; Tel.: +1-514-848-2712 ext. 2712}

\firstnote{Current address: 1455 De Maisonneuve Blvd. W. EV-009.187, Montreal, Quebec, Canada, HG-1M8}

\abstract{
Recent advancements in Electroencephalography (EEG) sensor technologies and signal processing algorithms have paved the way for further evolution of Brain Computer Interfaces (BCI) in several practical applications ranging from rehabilitation systems to smart consumer technologies. When it comes to Signal Processing (SP) for BCI, there has been a surge of interest on Steady-State motion-Visual Evoked Potentials (SSmVEP), where motion stimulation is utilized to address key issues associated with  conventional light-flashing/flickering. Such benefits, however, come with the price of having less accuracy and less Information Transfer Rate (ITR). In this regard, the paper focuses on the design of a novel SSmVEP paradigm without using resources such as trial time, phase, and/or number of targets to enhance the ITR. The proposed design is based on the intuitively pleasing idea of integrating more than one motion within a single SSmVEP target stimuli, simultaneously. To elicit SSmVEP, we designed a novel and innovative dual frequency aggregated modulation paradigm, referred to as the  Dual Frequency Aggregated steady-state motion Visual Evoked Potential (DF-SSmVEP), by concurrently integrating ``Radial Zoom'' and ``Rotation'' motions in a single target without increasing the trial length. Compared to conventional SSmVEPs, the proposed DF-SSmVEP framework consists of two motion modes integrated and shown simultaneously each modulated by a specific target frequency. The paper also develops a specific unsupervised classification model, referred to as the Bifold  Canonical Correlation Analysis (BCCA), based on two motion frequencies per target. The corresponding covariance coefficients are utilized as extra features improving the classification accuracy. The proposed DF-SSmVEP is evaluated based on a real EEG dataset and the results corroborate its superiority. The proposed DF-SSmVEP outperforms its counterparts and achieved an average ITR of 30.7$\pm$1.97 and an average accuracy of 92.5$\pm$2.04, while the Radial Zoom and Rotation result in average ITRs of 18.35 $\pm$ 1 and 20.52 $\pm$2.5, and average accuracies of 68.12 $\pm$ 3.5 and 77.5$\pm$3.5 respectively.}


\begin{document}

\section{Introduction}
Recent advancements in Electroencephalography (EEG) sensor technologies and Signal Processing (SP) algorithms have paved the way for further evolution of Brain Computer Interface (BCI) systems~\cite{Jane:2019}. The ultimate goal of a BCI system is to establish a robust communication channel with high throughput and accuracy between brain and the outer world. BCI systems have found~several practical applications ranging from rehabilitation/assistive systems to diagnosis/prognosies of neurological disorders~\cite{Soroosh:2018, samanta, dagois}. Recent technology trends show that leading technology companies are racing to develop advanced  BCI systems coupled with Augmented Reality (AR) visors. It is widely expected that AR coupled with BCI would be the next era of computing. In this regard and for integration within an AR environment,

\vspace{.1in}
\noindent
\textbf{Literature Review}:
Generally speaking, there are two main visual BCI Paradigms, (1) \textit{Steady-State Visually Evoked Potential (SSVEP)~\cite{Kubacki:2021, Ikeda:2021, Guevara:2021, Chen:2021, zhang2019hierarchical, zhao2017ssvep,  nakanishi2014high, zhang2012multiple, wei2016stimulus, Kadioglu}}, where light-flashing (flickering) visual stimulus is used to induce evoked potentials in the EEG signals, and;
(2) \textit{Steady-State motion-Visual Evoked Potentials (SSmVEP)~\cite{Zhang:2021, yan2019steady, han2018highly, chai2019radial, beveridge2019neurogaming}}, where instead of using flickering, some form of graphical motion is used to evoke potentials.
The former category (SSVEP) has been the main research theme due to its high achievable Information Transfer Rate (ITR), minimal requirement for user training, and excellent interactive potentials, such as high tolerance to artifacts and robust performance across users. However, flickering light, causes extensive mental stress. Continuous use of SSVEPs (looking at flickering patterns for a long period of time), therefore, may cause seizure or eye fatigue.
The second category (SSmVEP) is introduced to address these issues while keeping all the aforementioned benefits of the SSVEPs.
\begin{figure}[t!]
~~~~~~~~~~~~~~~~~~~~~~~~\includegraphics[scale=0.45]{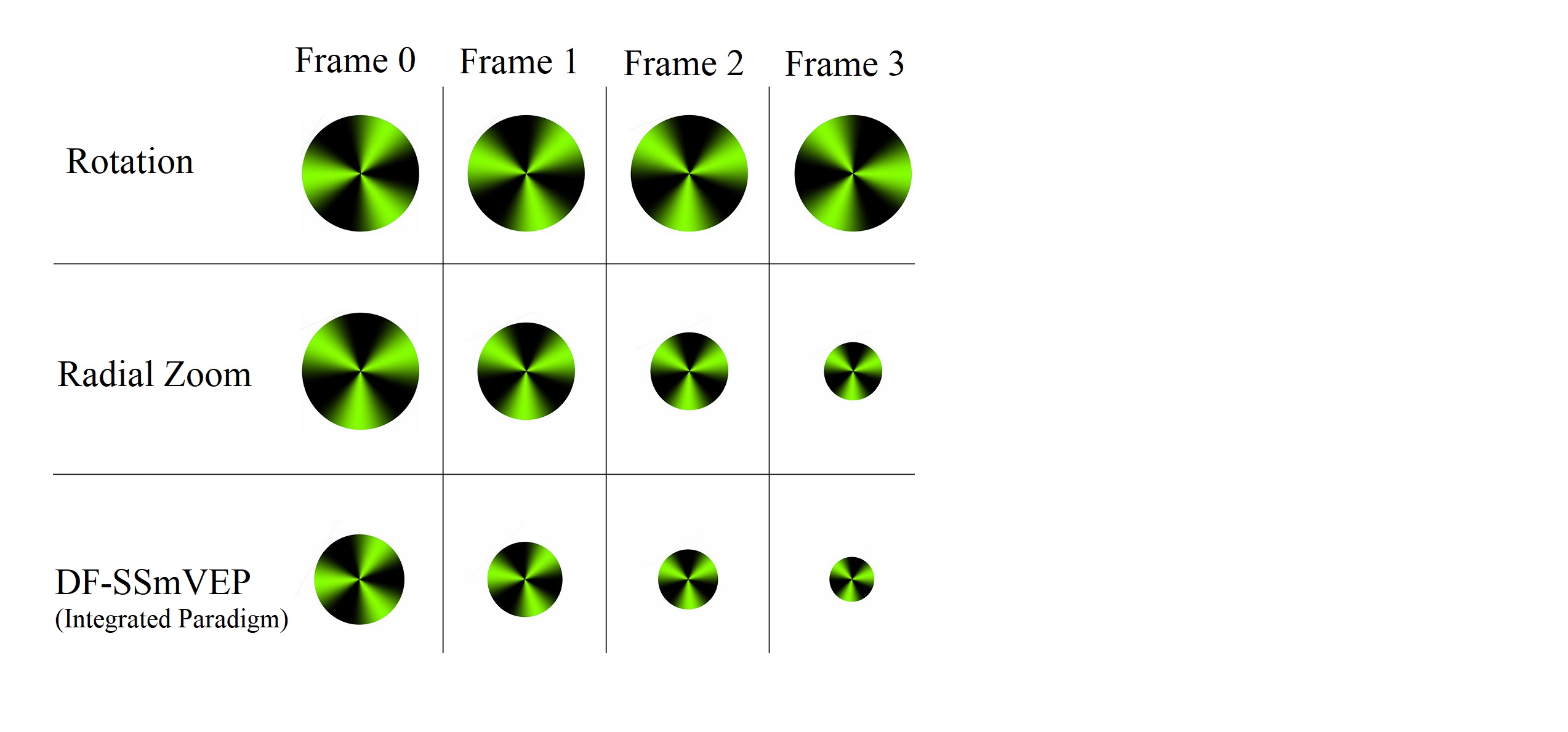}
\caption{\footnotesize Proposed $\DFS$ paradigm  developed by concurrent inclusion of two types of the motion (rotation and resizing).}
\label{Fig:DF-SSmVEP}
\end{figure}

One issue that is shared by both categories is the challenge of coding more targets under available resources. Within the context of SSVEPs, the following research works have been conducted to address this issue: Reference~\cite{nakanishi2014high} introduced simultaneous phase and frequency modulations. Reference~\cite{zhang2012multiple} used modulation in time, i.e.,  lengthening of the trial duration. Two different flickering-frequencies are shown consecutively for each target. Following a similar path, References~\cite{zhao2017ssvep} focused on using more than one frequency in a single target (but not at the same time) via Frequency-Shift Keying (FSK) modulation, also referred to as code modulation, i.e., trial time is again used for modulation purposes. Similarly, Reference~\cite{wei2016stimulus} used code modulation with a single frequency for each target but with different phase shifts over one trial to enhance the system. Such  code-VEPs~\cite{Kadioglu, zhao2017ssvep, wei2016stimulus} and phase modulation~\cite{nakanishi2014high, wei2016stimulus} techniques are, however, very sensitive to synchronization, and as trial time is used for modulation purposes, ITR will be compromised. When it comes to SSmVEPs, the issue of coding more targets with enhanced ITR has not yet been considered, the paper addresses this~gap.

\vspace{.1in}
\noindent
\textbf{Contributions}:
To address the aforementioned problem, we focus on designing a novel SSmVEP paradigm without using additional  resources such as trial time, phase, and/or number of targets to enhance the ITR. The proposed design is based on the intuitively pleasing idea of using more than one simultaneous motion within a single SSmVEP target stimuli. More specifically, as shown in Fig.~\ref{Fig:DF-SSmVEP}, to elicit SSmVEP we designed a novel and innovative dual frequency aggregated modulation paradigm, referred to as the $\DFS$, by concurrently integrating ``Radial Zoom'' and ``Rotation'' motions in a single target without increasing the trial length. Fig.~\ref{Fig:DF-SSmVEP2}(i) visually compares four different paradigms:  Conventional SSVEP frequency modulation is shown in Sub-figure (a), where 2 target frequencies, ``F1'' and ``F2'', are evoked in 2 different trials via flickering. Sub-figure (b) is similar to  Sub-figure (a) where now 2 target frequencies are used together, one after another by increasing the trial time. Sub-figure (c) in Fig.~\ref{Fig:DF-SSmVEP2}(i) illustrates 2 SSmVEP modulations similar to Sub-figure (a), but target frequencies are evoked now via motion of the circle.   Sub-figure~(d) shows the proposed $\DFS$ design where now 2 target frequencies are used together simultaneously eliminating the need to increase (sacrifice) the trial length for achieving higher accuracy as is the case in code/frequency modulated SSVEPs~\cite{zhao2017ssvep, wei2016stimulus}.

The paper also develops a specific unsupervised classification model adopted to the proposed innovative $\DFS$ paradigm. More specifically, in contrary to the existing works, we propose an unsupervised SSmVEP detection technique, referred to as the Bifold Canonical Correlation Analysis ($\DCCA$) utilizing unique characteristics of the proposed dual aggregated frequency design. The $\DCCA$ exploits  availability of two motion  frequencies for each target and separately considers each single frequency of the targets as a reference. The corresponding covariance coefficients are then used as extra features advancing the classification accuracy. The proposed $\DFS$ is evaluated based on a real EEG dataset.

\section{The Proposed $\DFS$}
\begin{figure}[t!]
\captionsetup[subfigure]{labelformat=empty}
\centering
\mbox{(i)\subfigure{\includegraphics[scale=0.5]{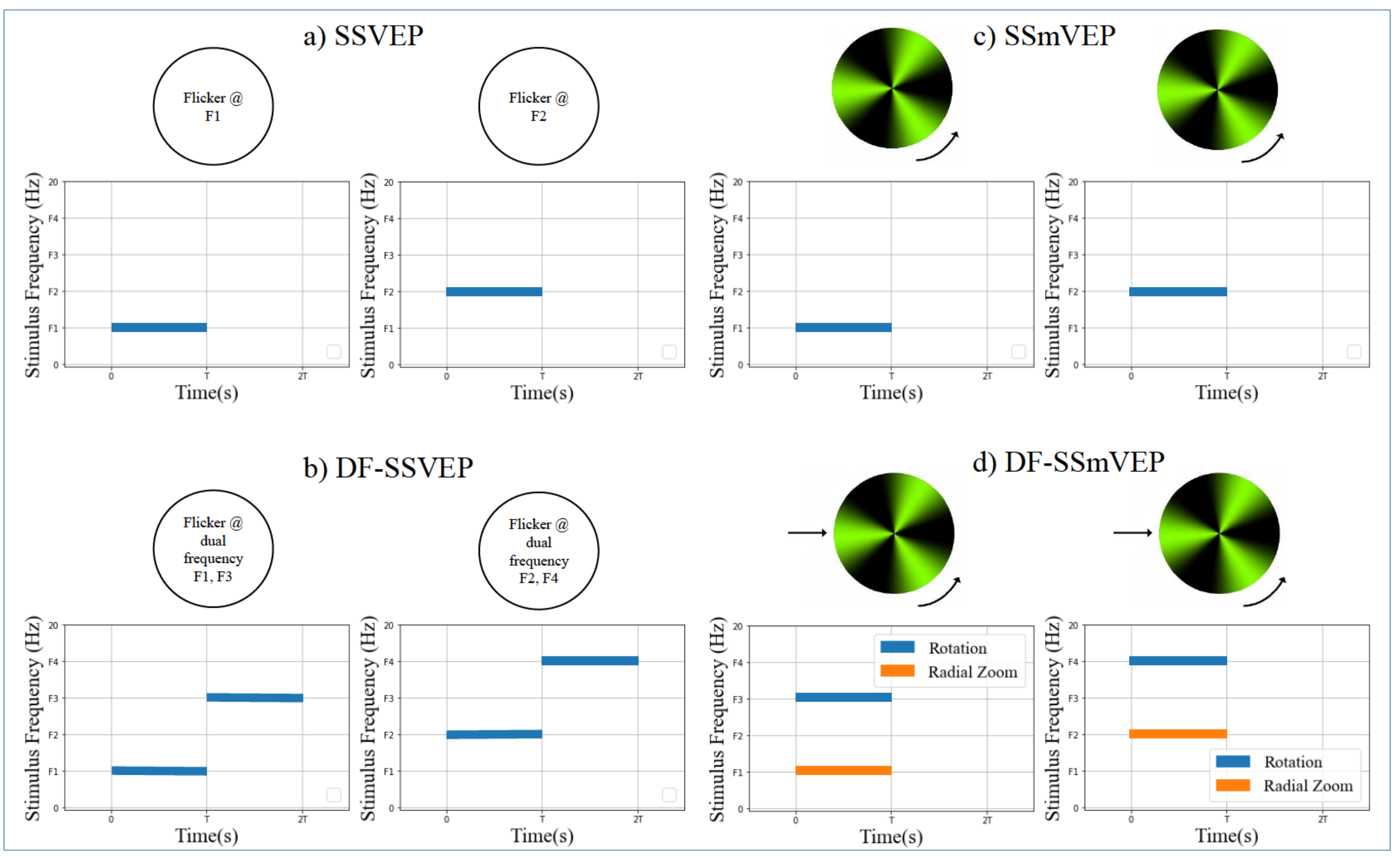}}}\\
\mbox{(ii)\subfigure{\includegraphics[scale=0.65]{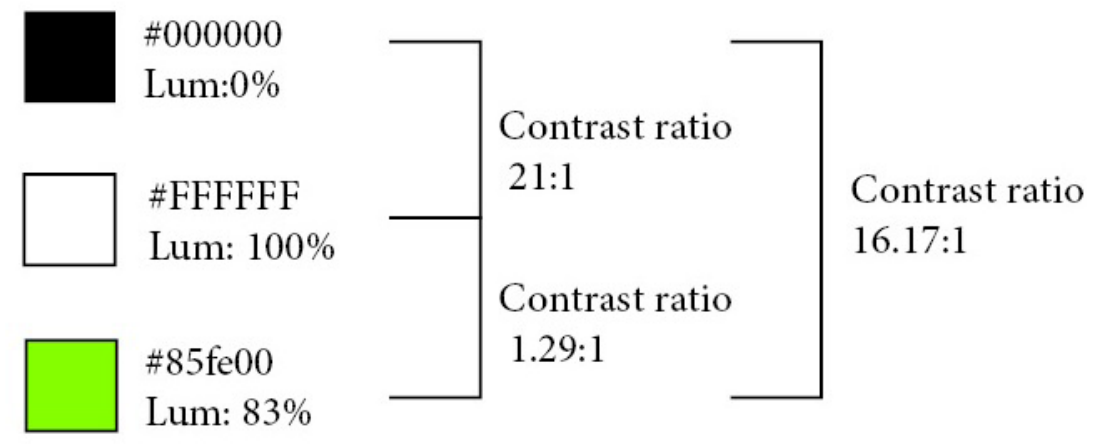}}}
\caption{\footnotesize  (i) Comparison between existing SSVEP frequency modulation schemes ((a) and (b)) with SSmVEP (c), and the proposed $\DFS$ (d). (ii) Luminance contrast ratio of the colors used in the designed $\DFS$. \label{Fig:DF-SSmVEP2}}
\end{figure}

In this section, we present the proposed $\DFS$ framework. The designed $\DFS$ stimulation paradigm includes a green and black circle with two motion modes. The first motion is the ``Radial Zoom'' in which the size of the circle changes periodically. The second mode is the ``Reciprocal Rotation'' of the circle between  $-45$$^\circ$ and $75$$^\circ$. Radial zoom motion and rotation motion are selected as candidates for integration following previous evaluations~\cite{yan2017four, chai2019radial}. The frequency of motion direction change inside the reciprocal motion is defined as motion inversion frequency. Furthermore, the motion inversion frequency corresponds to the frequency of stimulation, which is equal to the fundamental SSmVEP frequency. To elaborate on the motion choices utilized to design the proposed $\DFS$ paradigm, first we note that as mentioned in~\cite{yan2017four}, any paradigms with periodic motion can be used as stimuli of SSmVEPs. The two designs are integrated such that the focal point of one paradigm is overlaid with that  of the second one. In the proposed $\DFS$ paradigm, the focal point will be the center of the black-green circle, i.e., the center of oscillation for the two segments of the design (resizing and oscillation of circle).  To make the proposed design as efficient as possible and to  reduce fatigue~\cite{nakanishi2013approximation}, the proposed $\DFS$ design does not include high contrast colors improving the practical applicability of the SSmVEP stimuli.  As shown in Fig.~\ref{Fig:DF-SSmVEP2}(b), the Luminance Contrast Ratio (LCR) associated with our green-black paradigm is lower than that of the conventional black-white design.

The refresh rate of the monitor showing the SSmVEP stimuli is a limiting factor restricting the frequencies that can be designed when an equal number of frames used during consecutive cycles (one motion direction change). To implement flexible target frequencies, we need to have designs with variable number of frames per cycle~\cite{nakanishi2013approximation} to implement our design. Given the refresh rate of the monitor, which is $60$ Hz in our setting, the first task is to design binary stimulus sequences (i.e., the number of frames per each half-cycle, typically, asymmetric) with the goal of allocating frames based on the specified target frequencies. The number of frames per half-cycle~\cite{nakanishi2013approximation} of the $\DFS$ is constructed as $S(f\Tar,i) = \text{square}\big[2 \pi f\Tar(\dfrac{i}{R_{r}})\big]$,
where $i$ indicates the frame index; $f$ denotes the target frequency; $R_{r}$ represents the monitor's refresh rate, and; $S(f,i)$ denotes the stimulus sequences associated with target frequency $f$. Note that, each half-cycle corresponds to one contraction or expansion or half of the reciprocal motion. Consequently,  a motion oscillation at a target SSmVEP frequency up to $f \leq R_{r}/k$ can be generated as a stimulus. In other words, for one cycle of an SSmVEP paradigm to be understandable, $k$ minimum number of  frames per half-cycle is required to realize the motion cycle comfortably.

\vspace{.1in}
\noindent
\textbf{\textit{Coding Algorithm}}:
Assume that the maximum number of targets (objects shown on the screen simultaneously) is  denoted by $\NT$. In other words, $\NT$ number of target frequencies are selected within the limited frequency spectrum of [$f_{\min}, f_{\max}$] available for constructing the stimuli.  Term $f_{i}$, for ($1 \leq i \leq \NT$), represents the target frequency for the $i^{\text{th}}$ target/object assumed to be sorted in an ascending order, i.e., ($f_{\min} \leq f_1 < f_2 < \ldots < f_{\NT} \leq f_{\max}$). These $\NT$ target frequencies need to be derived in an intelligent fashion such that the best performance among all the susceptible frequencies is achieved (i.e., achieve accuracy improvements without reducing the ITR).
\textit{In the proposed dual aggregated design, each target includes two motions with two distinct frequencies.} These frequencies are assigned to targets in which no two pairs of targets have more than one adjacent frequencies. More specifically, the objective of the coding algorithm is to find these two underlying target frequencies in such a way that each pair of objects at most have one adjacent target frequency. For each consecutive frequencies $f_{i}$ and $f_{i+1}$,  $g_{i}$ is defined~as
\begin{equation}
g_{i} =
\begin{cases}
\dfrac{f_{i} + f_{i+1}}{2},  \:\:\:\:\:               \forall i \in [1,N-1]\\
f_{i} + M,        \:\:\:\:\:\:\:\:\:\:   i = N
\end{cases}
 \end{equation}
where $M = \min[\dfrac{f_{i+1}-f_{i}}{2}]$ $\forall i \in [1,N-1]$. Each $g_{i}$ is adjacent to $f_{i}$ and $f_{i+1}$. More specifically, consider $P=\{(a_{i},b{i})\}$, for ($1 \leq i \leq N$), representing the set of $N \geq 5$ target pairs where $a_{i}$ and $b_{i}$ are the new SSmVEP frequencies used for the $i^{\text{th}}$ object. Terms $a_{i}$ and $b_{i}$ in $P$ are defined as follows
\begin{equation}
a_{i} =\begin{cases}
f_{1},    \:\:\:\:\:          i = 1\\
f_{i+1},   \:   \forall i \in [2,N-1] \\
f_{2},     \:\:\:\:\:          i = N
\end{cases}
b_{i} =\begin{cases}
g_{N-1},  \:\:\:\:\:           i = 1\\
g_{i-1},        \:\:\:\:\:\:   \forall i \in [2,N-1] \\
g_{N},  \:\:\:\:\:\:\:\:\:\:             i = N
\end{cases}.
\end{equation}
%

\vspace{0.1in}
\noindent
\textbf{\textit{Pre-processing:}}
The proposed SSmVEP paradigm is implemented via a BCI system for real EEG data collection. In this regard, the first step is pre-processing of EEG signals associated with the proposed SSmVEP paradigm, as collected EEG signals are exposed to artifacts and high/low frequency noises. To extract the SSmVEP signal from the EEG signals, applying spatial and time domain filters are, therefore, critical. In time domain filtering, first, zero-phase Chebyshev Type I band-pass filter ($2$-$40$ Hz) is applied to smooth the data and remove high-frequency artifacts.

\subsection{Proposed $\DCCA$ Paradigm}

\noindent
Canonical Correlation Analysis (CCA) is a statistical method to study the linear relationship between two groups of multi-dimensional variables. For  two sets of signals arranged in matrices denoted by $\X$ and $\Y$, the goal is to find two linear projection vectors $\w_{x}$ and $\w_{y}$, such that the  linear combination of two groups of  signals $\w_{x}^{T}\X$ and  $\w_{y}^{T}\Y$ has the largest correlation coefficient, i.e.,
\begin{equation}
\rho = \max \frac{E(\w_{x}^{T}\X\Y^{T}\w_{y})}{\sqrt{E(\w_{x}^{T}\X\Y^{T}\w_{x})E(\w_{y}^{T}\X\Y^{T}\w_{y})}}.
\end{equation}
Reference signals are constructed at the stimulation frequency $f_{i}$~as
\begin{equation}
\y_{i}=[\cos 2\pi f_{i}t, \sin 2\pi f_{i}t, \ldots, \cos 2\pi \NH f_{i}t, \sin 2\pi \NH f_{i}t]^T,
\end{equation}
where $t = \frac{1}{f_{s}}, \ldots,\frac{m}{f_{s}}$, the $f_{s}$ is the sampling rate, $m$ is sample point, and $\NH$ is the number of harmonics, which is dependent on the paradigm and is  obtained experimentally from Welch Power Spectrum of signals. Consider $\X$ as the matrix of the EEG signals collected from $K$ different channels. The CCA finds linear combination of coefficients with the largest correlation between $\X$ and $\bm{Y}$.
 In the $\DCCA$ fusion, there is a feature vector for each sample concerning each target frequency. Contemplating the Power Spectral Density (PSD) of the EEG signal, collected during the aggregated paradigm,  gives insight that only one of the peaks is significant for some trials, i.e., one of the two modulated frequencies has more impact on the visual pathways of the brain. To capitalize this unique property and enhance the DF-SSmVEP, the following three references are incorporated to create the feature~vector
\begin{eqnarray}
\y_{1}&=&  [\cos(2\pi f_{i,1}t), \sin(2\pi f_{i,1}t), \ldots,\cos(2\pi \NH f_{i,1}t), \sin(2\pi \NH f_{i,1}t)]^T\\
\y_{2}&=& [\cos(2\pi f_{i,2}t), \sin(2\pi f_{i,2}t), \ldots, \cos(2\pi \NH f_{i,2}t), \sin(2\pi \NH f_{i,2}t)]^T\\
\y_{c}&=& C(\y_{1}, \y_{2},[ \cos(2\pi (f_{i,1}+f_{i,2})t), \sin(2\pi (f_{i,1}+f_{i,2})t)]^T,2)\label{nEq:eq7}
\end{eqnarray}
where $f_{i,j}$ represents the $j^{\text{th}}$ stimulation frequency of $i^{\text{th}}$ target, and the operator $C(a,b,c,2)$ concatenates three matrices a, b , and c vertically. The projection of each vector is separately calculated leading to three different weight vectors between test signal $\X$ and: (i) ($w_{y_1},w_{X_1}$) sine/cosine reference of first frequencies; (ii) ($w_{y_2},w_{X_2})$  reference of second frequency of targets, and; (iii) ($w_{y_{c}},w_{X_{c}}$) sine/cosine reference of both frequencies of targets. The feature vector $\bm{v}$ is
\begin{eqnarray}
\bm{v} = \big[\rho_1, \rho_2, \rho_{c}\big]^T, \text{ and }
\rho_a =\dfrac{\rho_1 + \rho_2 + \rho_{c}}{3}, \label{eq:equation2}
\end{eqnarray}
where $\rho_a$ is used as the final value to represent the correlation between the unknown sample and the frequencies of a target. It is worth mentioning that the proposed $\DCCA$  is an unsupervised technique as such there is no need for a separate training step. Consequently, all the available trials of sessions are used in the testing stage.

\subsection{Experimental Setup}
\begin{figure}[t!]
\centering
\mbox{\subfigure[]{\includegraphics[scale=0.53]{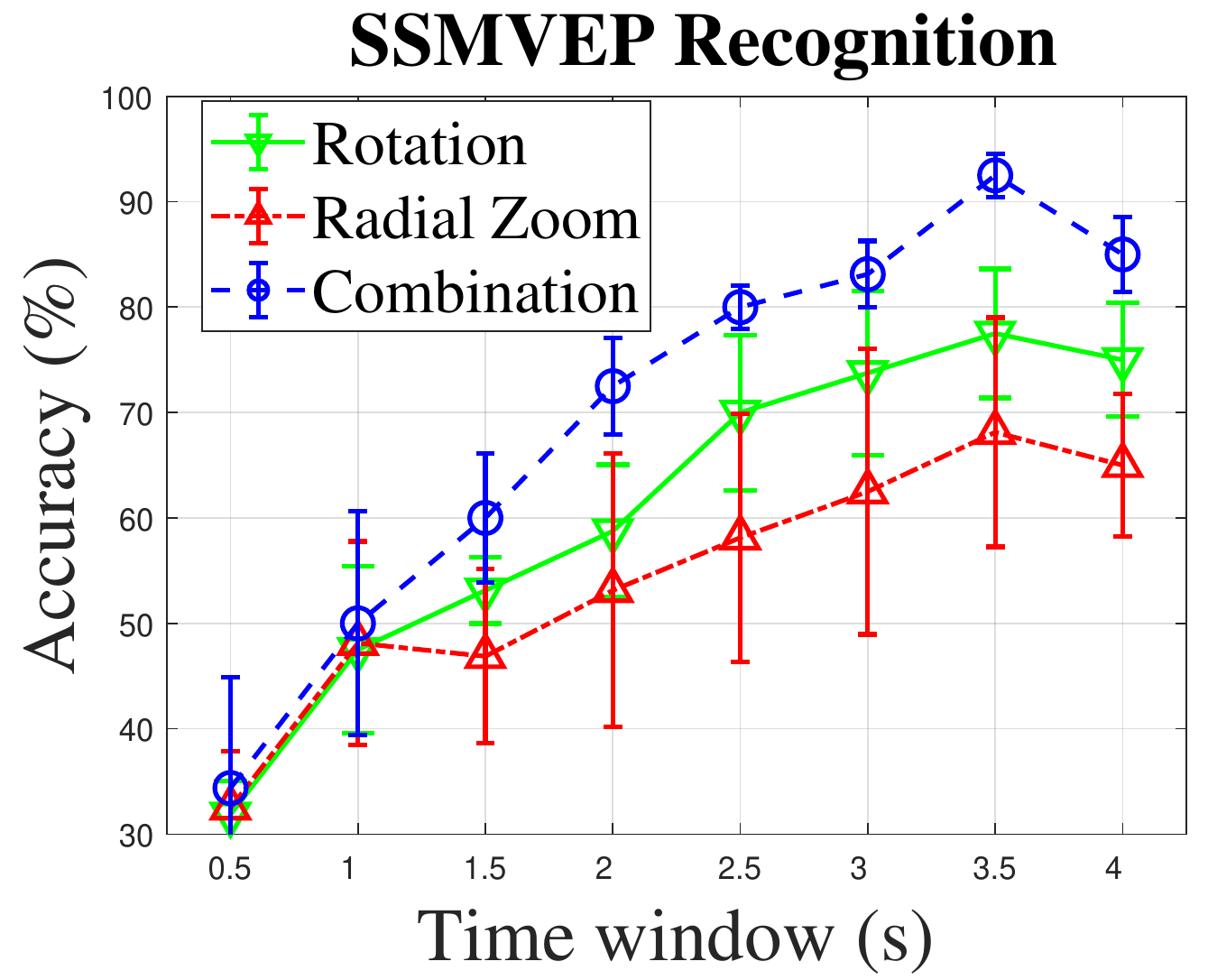}}
\subfigure[]{\includegraphics[scale=0.49]{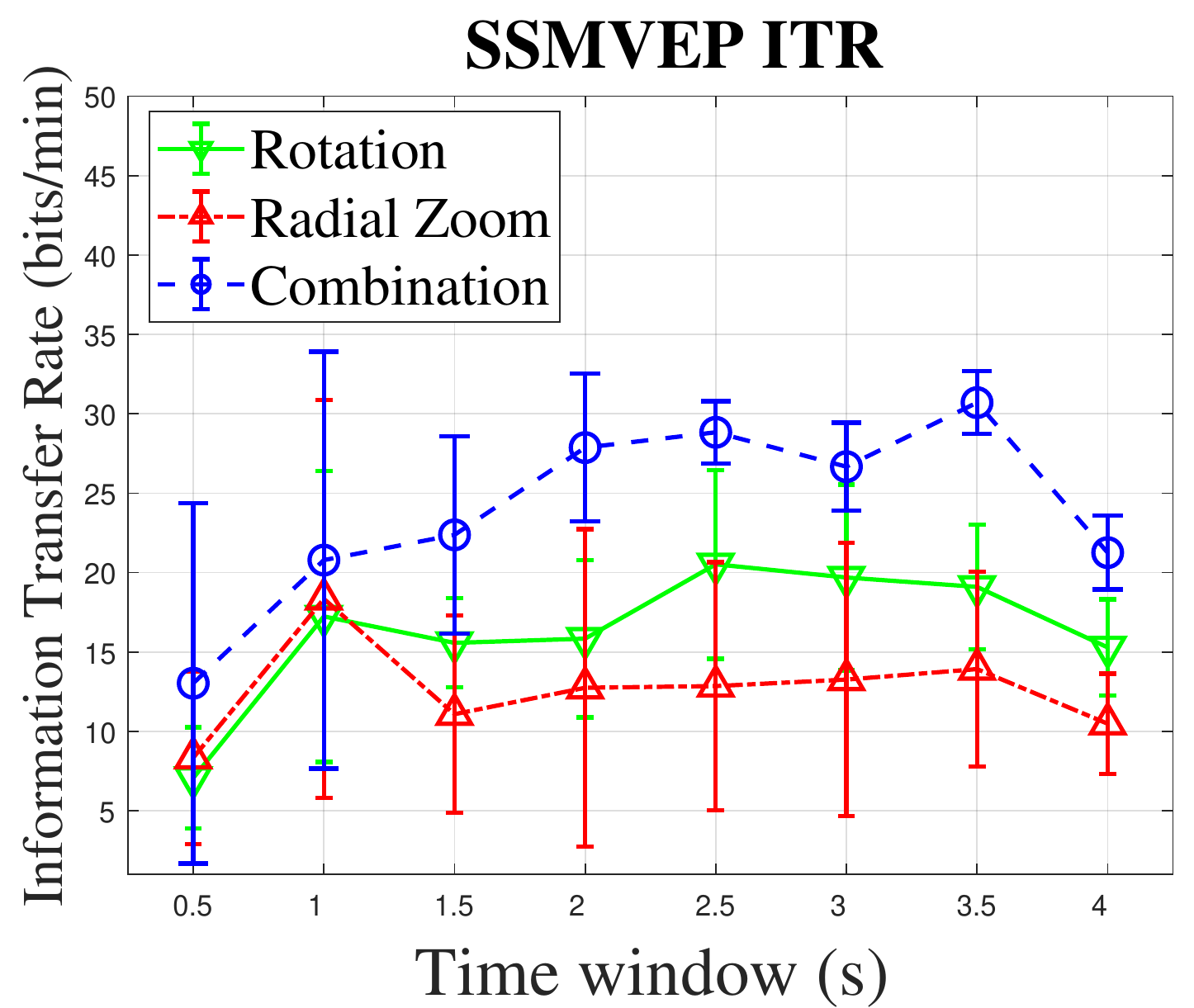}}}
\caption{\footnotesize Mean and standard deviation across all subjects for each time window: (a) Accuracy comparisons. (b) ITR comparisons.  \label{Fig:3}}
\end{figure}
\begin{figure}[t!]
\centering
\mbox{\subfigure[]{\includegraphics[scale=0.49]{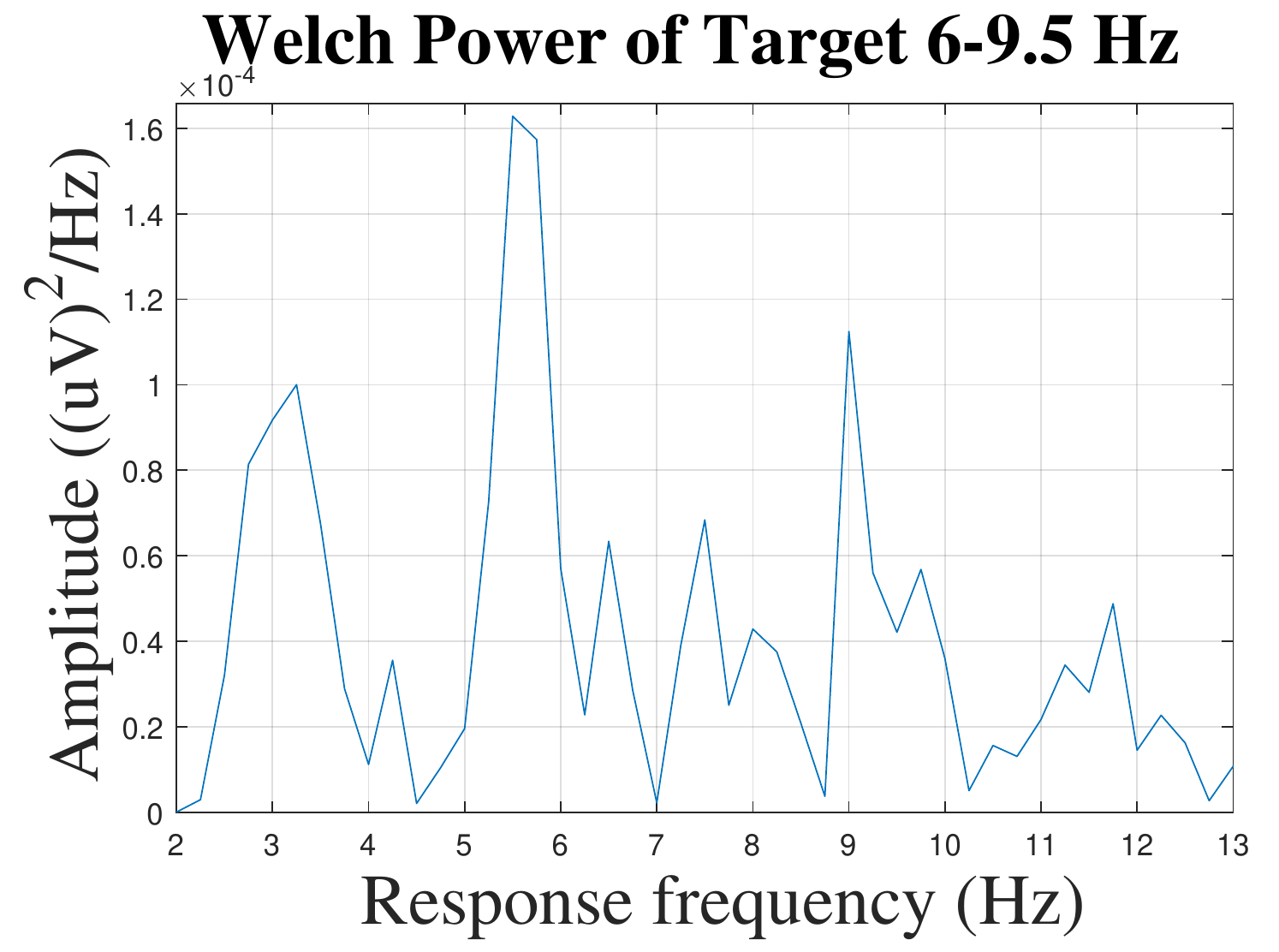}}
\subfigure[]{\includegraphics[scale=0.49]{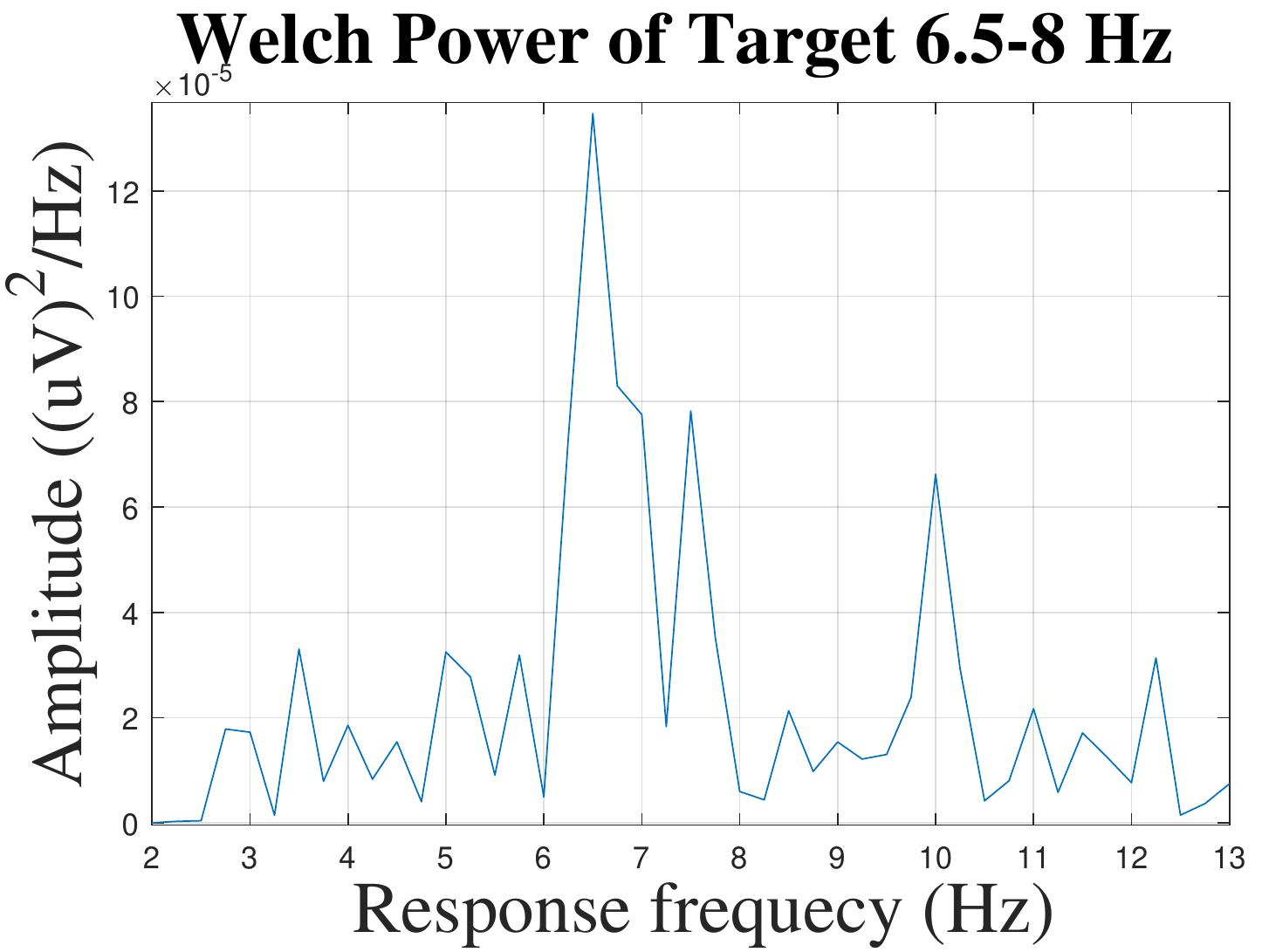}}}
\caption{\footnotesize (a) The PSD plot based on aggregated SSmVEP with $6$-$9.5$Hz target frequencies. (b) Similar to (a) but with $8$-$6.5$Hz target frequencies.  \label{Fig:PSD}}
\end{figure}
\begin{table}[t!]
\centering
\caption{\footnotesize Mean accuracy (\%) and mean ITR (bits/min) comparison between four methods of spatial filtering across the three motions. The best ITR among different time-windows are reported for each filter. Maximum Contrast Fusion (MCF)~\cite{yan2017four} and T-F Image Fusion~\cite{yan2019steady} are two types of spatial filtering.}
\label{table:1}
\begin{tabular}{|c | c | c | c | c | c | c |}
\hline
\backslashbox{\textbf{Filters}}{\textbf{Paradigms}} &\multicolumn{2}{c|}{\textbf{Radial Zoom}} &
\multicolumn{2}{c|}{\textbf{Rotation}} & \multicolumn{2}{c|}{\textbf{$\DFS$}}\\[0.5ex]
\cline{2-7}
&ACC & ITR & ACC & ITR & ACC & ITR\\
\hline
MCF + CCA  & \textbf{68.12}&\textbf{18.35} & \textbf{77.5}&\textbf{20.52} & 81.88 & 21.89 \\
\hline
T-F Image Fusion + CCA & 59.3 & 13.39 & 68.75 & 13.73 & 63.25&13.93\\
\hline
CCA Fusion & 63.5&17.24 & 76.17 &18.05& 84.38&23.73\\
\hline
~$\DCCA$ Fusion & - & - &-&-&\textbf{92.5}&\textbf{30.7}\\
\hline
\end{tabular}
\end{table}
\begin{sidewaystable} 
\centering
\caption{\footnotesize Comparison between mean and standard deviation values associated with performance indices (precision, sensitivity, specificity, and accuracy across) all runs for $3$ paradigms, i.e., Rotation (R), Radial Zoom (RZ), and the proposed $\DFS$ (denoted by DF).\label{table:2}}
\begin{tabular}{|c|c|c|c|c|c|c|c|c|c|c|c|c|}
\hline
 \backslashbox{\textbf{Classes}}{\textbf{PI}} &\multicolumn{3}{c|}{\textbf{Specificity}} &
\multicolumn{3}{c|}{\textbf{Sensitivity}} & \multicolumn{3}{c|}{\textbf{Precision}} &
\multicolumn{3}{c|}{\textbf{Accuracy}}\\
\cline{2-13}
&R  &RZ & $DF$ &R& RZ &DF  &R &RZ &DF &R&RZ&DF \\
\hline
9 Hz or&0.912&0.934&\textbf{0.993}&\textbf{0.900}&0.600&0.875&0.735&0.710&\textbf{0.977}&0.910&0.867&\textbf{0.970}\\ (9Hz, 7.5Hz)  & $\pm$ 0.052 & $\pm$ 0.037 & $\pm$ \textbf{0.013} & $\pm$ 0.098 & $\pm$ 0.226  &$\pm$ \textbf{0.083}&$\pm$0.137&$\pm$ 0.061& $\pm$ \textbf{0.047}&$\pm$ 0.054&$\pm$ 0.020& $\pm$ \textbf{0.01} \\
\hline
6 Hz or&0.940&0.893 &\textbf{0.984}&0.725&0.875&\textbf{0.962}&0.757&0.702&\textbf{0.947}&0.897&0.890&\textbf{0.980}\\ (6Hz, 9.5Hz)& $\pm$ \textbf{0.023} & $\pm$ 0.062 &$\pm$ 0.027  &$\pm$ 0.098&$\pm$ \textbf{0}&$\pm$ 0.060&$\pm$\textbf{0.091}&$\pm$ 0.167&$\pm$ 0.870&$\pm$ 0.027&$\pm$ 0.050&$\pm$ \textbf{0.023} \\
\hline
5 Hz or&\textbf{0.946}& 0.875 &0.940&0.625&0.937&\textbf{0.975}&0.737&0.663&\textbf{0.829}&0.882&0.8875&\textbf{0.952}\\ (5Hz, 8.5Hz)&$\pm$ \textbf{0.025} & $\pm$0.044&$\pm$0.033  &$\pm$0.220& $\pm$ 0.106&$\pm$\textbf{0.053}&$\pm$0.118&$\pm$ \textbf{0.068}&$\pm$0.100&$\pm$ 0.054&$\pm$ \textbf{0.017}&$\pm$0.029 \\
\hline
7 Hz or&0.981& 0.993&\textbf{0.993}&0.787&0.312&\textbf{1}&0.937&0.933&\textbf{0.977}&0.942&0.857&\textbf{0.995}\\ (7Hz, 5.5Hz)& $\pm$ 0.030  & $\pm$ 0.013&$\pm$ \textbf{0.013}  &$\pm$ 0.177&$\pm$ 0.135&$\pm$\textbf{0}&$\pm$0.100&$\pm$ 0.140&$\pm$\textbf{0.047}&$\pm$ 0.026&$\pm$ 0.031&$\pm$\textbf{0.011} \\
\hline
8 Hz or&0.940&0.906&\textbf{0.987}&\textbf{0.850}&0.687&0.812&0.809&0.649&\textbf{0.939}&0.922&0.862&\textbf{0.952}\\ (8Hz, 6.5Hz)& $\pm$0.047 &$\pm$ 0.062 &$\pm$\textbf{0.016}&$\pm$ \textbf{0.098}&$\pm$ 0.244&$\pm$0.135&$\pm$0.131&$\pm$ 0.221&$\pm$\textbf{0.079}&$\pm$ \textbf{0.027}&$\pm$ 0.095&$\pm$0.036 \\
\hline
\end{tabular}
\end{sidewaystable}
A real dataset consisting of $10$ individuals, $5$ women and $5$ men between age of $20$ to $27$  is utilized to evaluate the proposed $\DFS$ framework. Participants have no evidence of visual or color-recognition ailments. Five subjects had experience of BCI experiments.
We would like to mention that, for data collection, we have followed the common, standard, and accepted approach in the BCI domain (e.g.,~\cite{yan2017four}) where, typically $9$-$11$ subjects are used, and the trial length ranges from $2$ seconds to $6$ seconds.
 The EEG signals were collected using a portable and wireless bio-signal acquisition system ($32$ bipolar active wet electrodes with sampling rate $500$Hz), g.Nautilus from g.tech Medical Engineering. The reference and the ground electrodes of the headset were placed at the earlobe and frontal position (Fpz), respectively. The electrodes $P_{z}$, $P_{o7}$, $P_{o3}$, $P_{o4}$, $P_{o8}$, and $O_{z}$ from the parietal-occipital region were chosen to collect  EEG signals. Stimuli with a white background were displayed on a $21.5$-inch LED screen at $60$Hz refresh rate. The resolution of the screen was $1920 \times 1080$, and the viewing distance was $70$cm. The data were collected with the policy certification number $3007997$ of Ethical acceptability for research involving human subjects approved by Concordia University.

\vspace{.1in}
\noindent
\textbf{\textit{Experimental Protocol}}:
Three  paradigms, i.e., (i) Rotation; (ii) Radial zoom, and; (iii)  $\DFS$, are tested in separated runs (videos are available publicly at~\cite{RaykaVid}). Each video consists of five targets oscillating with different frequencies. For individual videos, target frequencies were $5$, $6$, $7$, $8$, and $9$Hz. For aggregated videos, target frequencies of the radial zoom pattern were $5$, $6$, $7$, $8$, and $9$Hz, and target frequencies of the rotation pattern were $5.5$, $6.5$, $7.5$, $8.5$, and $9.5$Hz.  Each run of a video included two consecutive sessions, where subjects were required to stare at a target using a pointer.  In each session, each existing target was pointed in four trials. Each trial lasted $3.5$ seconds with a $2.5$ seconds break between consecutive trials. In particular, ITR is used for evaluations, which assesses the speed of a BCI systems as
\begin{eqnarray}
\text{ITR} =\frac{60}{T}\left[\log_2K +\sigma\log_2\sigma +(1-\sigma)\log_2(\frac{1-\sigma}{K-1})\right],
\end{eqnarray}
where $T$ is the sum of time of each trial and the resting state time between two trials; $K$ is the number of stimuli, and; $\sigma$ is the recognition accuracy. Evaluation via the One-way Analysis Of Variance (ANOVA)~\cite{xie2012steady} with Tukey post hoc analysis is also used to confirm that responses to the proposed $\DFS$-stimuli is statistically meaningful ($p < 0.05$). Some of the functions of ~\cite{zhang2014frequency} are utilized in the pipeline code.

\section{Result}
As shown in Fig.~\ref{Fig:3}, accuracy and ITR are measured for different time windows for each trial ranging from $0.5$ to $4$ seconds with an interval of $500$ millisecond. The highest rate of transmission (ITR) belongs to aggregated motion ($30.7\pm1.97$), which also achieved the best accuracy ($92.5\pm2.04$). We utilized the ANOVA test  following the existing literature~\cite{yan2017four,han2018highly} that used ANOVA test for significance comparison between accuracies and/or ITRs of different paradigms. It is worth noting that based on the Central Limit Theorem, we can safely assume that the samples have a normal distribution.  The one-way ANOVA on accuracies of $\DFS$ paradigm reveals that there is no significant effect of frequencies (classes) on accuracies (F = 2.78, p = 0.065 ), so all target frequencies of $\DFS$ are feasible in BCI systems. The Tukey post-hoc test on the accuracy and the ITR shows significant differences  between the performance of $\DFS$ with $\DCCA$ and other paradigms with corresponding classifiers for which the highest accuracies are acquired, i.e., (Accuracy: $p_{DF-R} = 0.042$, $p_{DF-RZ}=0.002$, $p_{R-RZ} = 0.218$; ITR: $p_{DF-R} = 0.011$, $p_{DF-RZ}=0.001$, $p_{R-RZ} = 0.264$).

Fig.~\ref{Fig:PSD} illustrates Welch PSD of two targets of aggregated motion after spatial filtering for Subject 1. Two significant peaks of SSmVEP frequency around the two   frequencies of each target is observable. Table~\ref{table:1} shows the overall recognition accuracies and ITRs. It can be observed that aggregating  two SSmVEP paradigms using the proposed $\DFS$ results in compelling performance improvement. The results also show superiority of the $\DCCA$ as the best unsupervised target detector among its counterparts. Means and standard deviations based on four different Performance Indices (PI), i.e., precision, sensitivity, specificity, and accuracy across all runs for each class, are also shown in Table~\ref{table:2}. The time window is set to $3.5$ second. Each row of Table~\ref{table:2} corresponds to the dual frequency of the proposed $\DFS$ paradigm in each class. To investigate robustness of the proposed methodology, we display paradigms repeatedly to different subjects (in contrary to using EEG signals of one trial several times) and average the performance across Runs and Subjects. In the experiments, we have $10$ subjects each performing (repeating) one run for $8$ times (resulting in 80 trials per class).

\section{Conclusion}
To address lower accuracy and ITR of SSmVEP designs, the paper proposed an intuitively pleasing, novel, and innovative dual frequency aggregated modulation paradigm. Referred to as the $\DFS$, the novel design is constructed by concurrently integrating ``Radial Zoom'' and ``Rotation'' motions in a single target without increasing the trial length.  The paper also develops a specific unsupervised classification model, referred to as the~$\DCCA$, which utilizes availability of two motion frequencies per each target. The proposed DF-SSmVEP is evaluated via a real EEG dataset achieving average ITR of 30.07$\pm$1.97 and average accuracy of 92.5$\pm$2.04.



\begin{thebibliography}{10}

\bibitem{Jane:2019}
X. Chen, \textit{et al.},
\newblock ``Removal of Muscle Artifacts From the EEG: A Review and Recommendations,"
\newblock {\em IEEE Sensors Journal}, vol. 19, no. 14, pp. 5353-5368, 2019.

\bibitem{Soroosh:2018}
S. Shahtalebi and A. Mohammadi,
\newblock ``Bayesian Optimized Spectral Filters Coupled With Ternary ECOC for Single-Trial EEG Classification,"
\newblock {\em IEEE Trans. Neural Syst. Rehabil. Eng.}, vol. 26, no. 12, pp. 2249-2259,  2018.

\bibitem{samanta}
K. Samanta, S. Chatterjee, and R. Bose, \newblock ``Cross-Subject Motor Imagery Tasks EEG Signal Classification Employing Multiplex Weighted Visibility Graph and Deep Feature Extraction,'' \newblock {\em IEEE Sensors Letters}, vol. 4, no. 1, pp. 1-4, 2020.

\bibitem{dagois}
E. Dagois, A. Khalaf, E. Sejdic and M. Akcakaya,  \newblock ``Transfer Learning for a Multimodal Hybrid EEG-fTCD Brain–Computer Interface,'' \newblock {\em  IEEE Sensors Letters}, vol. 3, no. 1, pp. 1-4, 2019.

\bibitem{zhang2019hierarchical}
Y. Zhang, \textit{et al.},
\newblock ``Hierarchical Feature Fusion Framework for Frequency Recognition in SSVEP-based BCIs,''
\newblock {\em  Neural Networks},  vol. 119, pp. 1--9, 2019.

\bibitem{Kubacki:2021}
A. Kubacki,
\newblock ``Use of Force Feedback Device in a Hybrid Brain-Computer Interface Based on SSVEP, EOG and Eye Tracking for Sorting Items,''
\newblock {\em Sensors}, vol. 21, 7244, 2021.

\bibitem{Ikeda:2021}
A. Ikeda, Y. Washizawa,
\newblock ``Steady-State Visual Evoked Potential Classification Using Complex Valued Convolutional Neural Networks,''
\newblock {\em Sensors}, vol. 21, 5309, 2021.

\bibitem{Guevara:2021}
D. R. De la Cruz-Guevara, W. Alfonso-Morales,
\newblock ``Caicedo-Bravo, E. Solving the SSVEP Paradigm Using the Nonlinear Canonical Correlation Analysis Approach,''
\newblock {\em Sensors}, vol. 21, 5308, 2021.

\bibitem{Chen:2021}
Y.-J. Chen, P.-C. Chen, S.-C. Chen, C.-M. Wu,
\newblock ``Denoising Autoencoder-Based Feature Extraction to Robust SSVEP-Based BCIs,''
\newblock {\em Sensors}, vol. 21, 5019, 2021.


\bibitem{zhao2017ssvep}
X. Zhao, D. Zhao, X. Wang, X. Hou,
\newblock ``A SSVEP Stimuli Encoding Method Using Trinary Frequency-shift Keying Encoded SSVEP (TFSK-SSVEP),''
\newblock {\em  Frontiers in Human Neuroscience}, vol. 11, 2017.

\bibitem{wei2016stimulus}
Q. Wei, S. Feng, Z. Lu,
\newblock ``Stimulus Specificity of Brain-Computer Interfaces based on Code Modulation Visual Evoked Potentials,''
\newblock {\em PloS One}, vol. 11, no. 5, 2016.

\bibitem{nakanishi2014high}
M. Nakanishi, \textit{et al.},
\newblock ``A High-Speed Brain Speller using Steady-State Visual Evoked Potentials,''
\newblock {\em Int. J. Neural Systems}, vol. 24, no. 06, 2014.

\bibitem{zhang2012multiple}
Y. Zhang, \textit{et al.},
\newblock ``Multiple Frequencies Sequential Coding for SSVEP-based Brain-Computer Interface,''
\newblock {\em PloS One}, vol. 7, no. 3, 2012.

\bibitem{Kadioglu}
B. Kadioglu,  I. Yildiz, P. Closas, M. B. Fried-Oken and D. Erdogmus,
\newblock ``Robust Fusion of c-VEP and Gaze,'' \newblock {\em IEEE Sensors Letters}, vol. 3, no. 1, pp. 1-4, 2019.

\bibitem{Zhang:2021}
X. Zhang, W. Hou, W. Wu, L. Chen, N. Jiang,
\newblock ``Enhancing Detection of SSMVEP Induced by Action Observation Stimuli Based on Task-Related Component Analysis,''
\newblock {\em Sensors}, vol. 21, 5269,  2021.

\bibitem{beveridge2019neurogaming}
R. Beveridge, S. Wilson, M Callaghan, D. Coyle,
\newblock ``Neurogaming with Motion-onset Visual Evoked Potentials (MVEPs): Adults versus Teenagers,''
\newblock {\em IEEE Trans. Neural Syst. Rehabil. Eng.}, vol. 27, no. 4, pp. 572-581, 2019.

\bibitem{yan2019steady}
W. Yan, G. Xu, L. Chen, and X. Zheng,
\newblock ``Steady-state Motion Visual Evoked Potential (SSmVEP) Enhancement Method based on Time-Frequency Image Fusion,''
\newblock {\em Computational Intelligence and Neuroscience}, 2019.

\bibitem{han2018highly}
C. Han, G. Xu, J. Xie, C. Chen, S. Zhang,
\newblock ``Highly Interactive Brain-Computer Interface based on Flicker-Free Steady-State Motion Visual Evoked Potential,''
\newblock {\em Nature Scientific Reports}, vol. 8, no. 1, 2018.
%
\bibitem{chai2019radial}
X. Chai, \textit{et al.},
\newblock ``A Radial Zoom Motion-based Paradigm for Steady State Motion Visual Evoked Potentials,''
\newblock {\em Frontiers in Human Neuroscience}, vol. 13, 2019.
%
\bibitem{yan2017four}
W. Yan, \textit{et al.},
\newblock ``Four Novel Motion Paradigms based on Steady-state Motion Visual Evoked Potential,''
\newblock {\em IEEE Trans. Biomed. Eng.}, pp. 1696--1704, Aug. 2018.
%
\bibitem{nakanishi2013approximation}
M. Nakanishi , Y. Wang, Y. Mitsukura, T.P. Jung,
\newblock ``An Approximation Approach for Rendering Visual Flickers in SSVEP-based BCI using Monitor Refresh Rate,''  \newblock {\em IEEE Int. Con. Eng. Medicine \& Biology Society (EMBC)}, pp. 2176--2179, 2013.
%
\bibitem{xie2012steady}
J. Xie, G. Xu, J. Wang, F. Zhang, and Y. Zhang,
\newblock ``Steady-state Motion Visual Evoked Potentials Produced by Oscillating Newton's Rings: Implications for
Brain-computer Interfaces,''
\newblock {\em Plos one},   vol. 7, no. 6, pp. e39707, 2012.



\bibitem{zhang2014frequency}
Y. Zhang, G. Zhou, J. Jin, X. Wang, and A. Cichocki,
\newblock ``Frequency recognition in SSVEP-based BCI using multiset canonical correlation analysis,''
\newblock {\em International journal of neural systems},   vol. 24, no. 04, 2014.

\bibitem{RaykaVid}
https://github.com/raykakarimi/DF-SSmVEP-videos \\

\end{thebibliography}
\end{document}